\documentclass[twocolumn]{aastex631}
\usepackage[normalem]{ulem}
\usepackage[utf8]{inputenc}
\usepackage{amsmath}
\setcitestyle{sort,comma}
\usepackage{csvsimple}

\usepackage[version=4]{mhchem} % Formula subscripts using \ce{}
\usepackage{seqsplit}
\usepackage{upgreek}
\defcitealias{Cherubim2026}{C26}

\providecommand{\apjl}{ApJL}
\providecommand{\apjs}{ApJS}
\providecommand{\aap}{A\&Aj}

\published{November 26, 2026}
\begin{document}

\title {No Persistent Helium Absorption in LHS\,1140\,b: Four JWST/NIRISS SOSS Transits and Multiepoch Stellar He\,I Variability}

\newcommand{\planetarium}{Planétarium de Montréal, Espace pour la Vie, 4801 av. Pierre-de Coubertin, Montréal, Canada}

\newcommand{\umontreal}{Institut Trottier de recherche sur les exoplan\`etes, D\'epartement de Physique, Universit\'e de Montr\'eal, Montr\'eal, Qu\'ebec, Canada}

\newcommand{\omm}{Observatoire du Mont-M\'egantic, Qu\'ebec, Canada}

\newcommand{\ipag}{Universit\'{e} de Grenoble Alpes, CNRS, IPAG, 38000 Grenoble, France}

\newcommand{\geneve}{Observatoire de Gen\`eve, D\'epartement d’Astronomie, Universit\'e de Gen\`eve, Chemin Pegasi 51, 1290 Versoix, Switzerland}

\author[0000-0003-0854-3002]{Amélie Gressier}
\affil{\umontreal}
\email{amelie.gressier@umontreal.ca}

\author[0000-0001-9291-5555]{Charles Cadieux}
\affil{\geneve}

\author[0000-0001-5485-4675]{René Doyon}
\affil{\umontreal}
\affil{\omm}

\author[0000-0002-2195-735X]{Louis-Philippe Coulombe}
\affil{\planetarium}
\affil{\umontreal}
%\affil{\csa}

\author[0000-0002-1199-9759]{Romain Allart} 
\affil{\umontreal}
\affil{\ipag}

\author[0000-0003-3506-5667]{Etienne Artigau}
\affil{\umontreal}
\affil{\omm}

\author[0000-0002-7613-393X]{François Bouchy}
\affil{\geneve}

%\linenumbers

%Abstract shortened to 250 words
%TC:ignore
\begin{abstract}
The search for atmospheres on temperate terrestrial planets is important for understanding how these objects form and evolve and their potential habitability. Recent transit observations of LHS\,1140\,b, a temperate ($T_\mathrm{eq} = 226$\,K) planet straddling the radius valley ($R_\mathrm{p} \approx 1.7 \,R_\oplus$), with the WINERED high-resolution spectrograph yielded a detection of planetary atmospheric escape through the measurement of excess absorption ($1.24\%\pm0.23\%$) in the metastable helium triplet, although a later second visit resulted in a nondetection. We analyze four transits of LHS\,1140\,b and two of LHS\,1140\,c observed with JWST/NIRISS SOSS. We find no evidence of helium absorption in either planet, with all amplitudes consistent with zero within $1\sigma$. For LHS\,1140\,b, we derive $3\sigma$ upper limits of $0.72\%$--$1.21\%$. Individual visits disfavor the reported WINERED absorption at $2.6\sigma$--$3.7\sigma$, while their joint constraint disfavors a persistent signal at $4.5\sigma$. We find that the stellar He\,I line of LHS\,1140 varies substantially between NIRPS and WINERED epochs, with both its strength and fractional variability consistent with the behavior of other M dwarfs. If the high-resolution detection is indeed planetary, the rate of such atmospheric loss events must be relatively low ($f = 22_{-12}^{+17}\%$). Alternatively, stellar He\,I variability may contribute to the reported excess absorption. Additional high-resolution observations, both in and out of transit, are required to distinguish between these scenarios.
\end{abstract}
%TC:endignore
%\tableofcontents

\section{Introduction}\label{sec:intro}

The detection of atmospheres on temperate and warm terrestrial planets remains elusive. Despite the precision of the JWST, which is sensitive enough to probe the $\sim$20--30\,ppm features expected from secondary atmospheres, transmission spectroscopy studies of rocky exoplanets have mostly produced nondetection of atmospheric signatures, along with constraints on the range of possible atmosphere scenarios that can reproduce the observations \citep[e.g.,][]{LustigYaeger2023,Lim2023,Piaulet2025_T1d,Espinoza2025,Glidden2025}. In addition to transmission spectroscopy, thermal emission observations with JWST MIRI LRS and the 12 and 15\,$\mu$m photometric filters have revealed dayside brightness temperatures consistent with rocks devoid of substantial atmospheres for warm and temperate planets \citep[e.g.,][]{Greene2023,Zieba2023,Gillon2025}. 

While transmission spectroscopy observations of rocky planets generally focus on the detection of molecular species, the presence of an atmosphere can also be inferred through the detection of excess absorption from species escaping it. Particularly, the metastable triplet of helium at 10\,833\,\AA\, \citep{Seager2000}, which compared to UV tracers (e.g., Ly-$\alpha$) is not absorbed by the interstellar medium, has proven a powerful tool to study atmospheric escape in gas giant planets from low- and high-resolution both from space and from the ground-based observations \citep[e.g.,][]{Spake2018,Allart2018}. Although the JWST does not have sufficient spectral resolution to resolve the helium triplet, JWST/NIRISS SOSS observations have yielded multiple He\,I detections for gas giants \citep[e.g.,][]{Fu2022,Allart2025,Krishnamurthy2026} and even down to the sub-Neptune regime \citep{Ahrer2025}. 

Because the likelihood of secondary atmosphere retention is expected to be highly correlated with irradiation, as predicted by the cosmic shoreline hypothesis \citep{Zahnle2014}, rocky planets in the habitable zone of their stars present an important opportunity for atmospheric reconnaissance. One such target is LHS\,1140\,b, a $R_\mathrm{p} = 1.7$\,R$_{\oplus}$, $M_\mathrm{p} = 5.6$\,M$_{\oplus}$ super-Earth on a 24.7-day orbit around an M4.5V host star \citep{Dittmann2017}. The latest mass and radius measurements of LHS\,1140\,b are indicative of a density that is slightly lower than expected from a purely rocky composition \citep{Cadieux2023}, consistent with either a water world or mini-Neptune scenario. The water world interpretation was later further supported by NIRISS/SOSS and NIRSpec/G395H transmission spectroscopy observations, which revealed a lack of atmospheric features \citep{Cadieux_2024,Damiano2024}. LHS\,1140\,b is also ongoing further investigation as part of the rocky worlds DDT program \citep{Redfield2024}, and is scheduled for a total of nine secondary eclipse observations to probe its dayside brightness temperature from MIRI 15\,$\mu$m photometric observations.

Recently, \citet{Cherubim2026} (hereafter \citetalias{Cherubim2026}) presented the detection of excess helium absorption ($1.24\pm0.23\%$) from a transit observation of LHS\,1140\,b using the WINERED \citep{Ikeda2016, Ikeda2022} high-resolution spectrograph mounted on the 6.5-meter Magellan Clay telescope. A later visit of the planet using the same observing setup yielded a nondetection, with an upper limit on the excess absorption of 0.6\%, which was attributed to a temporal variability of the atmospheric escape. The detection of helium escape on LHS\,1140\,b is consistent with a helium world scenario \citep{Hu2015}, in which the hydrogen that made up the bulk of the primordial atmosphere composition was lost through escape-driven mass fractionation \citep{Cherubim2024}.

This Letter presents an analysis of the four transits of LHS\,1140\,b that have been observed with the NIRISS SOSS instrument \citep{Doyon2023,Albert2023} on the JWST, focusing on the helium triplet in search of potential excess absorption. We do not find evidence of atmospheric escape in any of the visits, which yield excess absorption upper limits that are below the level that was observed with WINERED. %Considering that one event of atmospheric escape at the level measured from the first WINERED visit is detected out of six transits, we constrain the rate of escape events to $f = 22_{-12}^{+17}\%$, indicating a relatively low frequency of significant atmospheric escape. We also present an analysis of He\,I variabilty from archival near-infrared high-resolution spctroscopy showing that LHS\,1140 shows significant variability. 
We present the observations and data reduction methodology in Section \ref{sec:obs}. The metastable helium analysis is described in Section \ref{sec:helium}, and we present and discuss the results in Section \ref{sec:results}. Finally, we summarize our findings and conclude in Section \ref{sec:conclusions}.

\section{Observations}\label{sec:obs}
For our analysis, we use five JWST/NIRISS SOSS observing visits of the LHS\,1140 system, covering four transits of planet b and two of planet c. The first two observations of LHS\,1140\,b were obtained as part of program DDT~6543 (P.I.: C. Cadieux \& R. Doyon). They started on UT 2023 December 1 and UT 2023 December 25, with the second transit occurring on December 26. This second visit also captured the transit of LHS\,1140\,c. Two additional transits of LHS\,1140\,b were observed on UT 2025 August 10 and UT 2026 July 23 as part of program GO~7073 (P.I.: J. Lustig-Yaeger). A second transit of planet c was acquired as part of the same program on UT 2025 July 26. 

All observations were acquired using the NIRISS SOSS mode with the \texttt{SUBSTRIP256} subarray and the \texttt{NISRAPID} readout pattern, covering the two diffraction orders ($0.6$--$2.8\,\upmu$m). Each DDT~6543 observation consist of 949 integrations with three groups per integration, for a total exposure time of $5.80$\,hr. The two GO~7073 observations of LHS\,1140\,b corresponds to 547 integrations with five groups per integration, for a total exposure time of $5.01$\,hr. The observation of LHS\,1140\,c comprised 450 integrations with five groups, resulting in a duration of $4.12$\,hr. As reported by \citet{Cadieux_2024}, the first DDT~6543 observation was affected by a target-acquisition failure caused by an incorrect proper motion epoch in the target definition. The target acquisition was obtained using blind guiding. The spectral traces were displaced by $-157$ pixels along the dispersion axis and $-12$ pixels along the cross dispersion axis compared to their nominal position. 

\begin{figure*}[htpb]
    \centering
        \includegraphics[width=\textwidth]{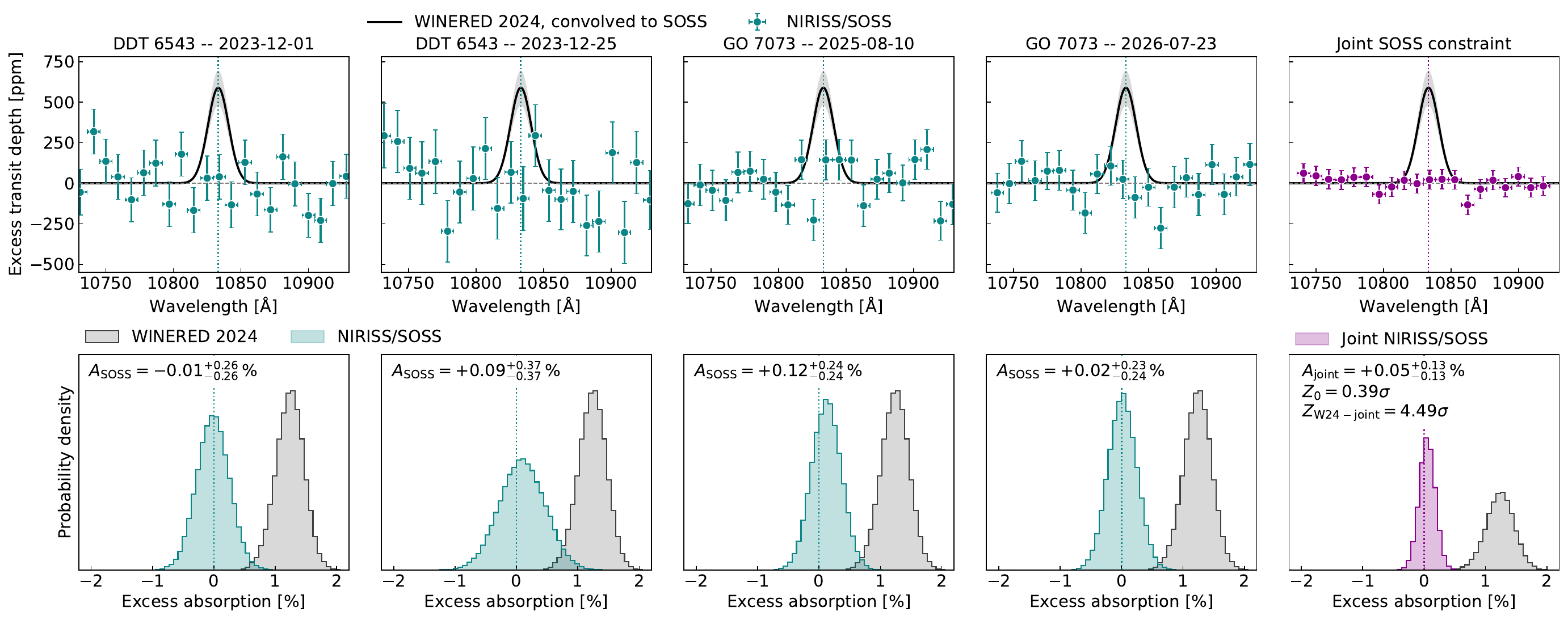}
    \caption{Results of the WINERED-informed helium analysis of LHS\,1140\,b. Top: continuum-subtracted NIRISS/SOSS spectra for the four visits (teal) and their inverse-variance-weighted mean (magenta) compared with the \citetalias{Cherubim2026} profile convolved to the SOSS resolution. Bottom: individual SOSS posteriors (teal) and the joint posterior from the four visits (magenta), compared with the \citetalias{Cherubim2026} posterior (gray). The joint posterior (right) shows a discrepancy with the WINERED measurement at 4.5$\sigma$. }
    \label{fig:helium_intrinsic_posteriors}
\end{figure*}

\subsection{Data Reduction}
We reduced the five NIRISS/SOSS visits from the \texttt{\_uncal.fits} files using the \texttt{transitspectroscopy} pipeline \citep{espinoza_2022}, following a procedure similar to those described in \citet{Gressier_2025} and \citet{Louie_2025}. The standard detector-level corrections included data-quality checks, saturation detection, superbias, reference-pixel, linearity, dark current, jump-detection, ramp-fitting and gain-scale corrections. We used the \texttt{transitspectroscopy}'s time-series-optimized  jump-detection procedure to produce one count-rate image per integration. 

The trace positions and the wavelength solution were obtained with \texttt{PASTASOSS} \citep{baines2023a, baines2023b}, except for the first DDT~6543 visit because the target-acquisition failure misplaced the spectral orders. For this visit, we located each order by cross-correlating the spatial profiles of the median detector image with a double-Gaussian model, measured the traces column by column and smoothed  them using piecewise spline functions. For order~1, we cross-correlated its continuum-normalized median spectrum with that of the second visit of DDT~6543 whose traces were correctly located and mapped its \texttt{PASTASOSS} wavelength solution using the 157 pixels displacement. This correction modified only the wavelength coordinates. For the background subtraction, $1/f$ noise and flux extraction we use the same procedure as described in \citet{Gressier_2025} and \citet{Louie_2025}. 

We extracted the spectra using a box aperture with a radius of 15-pixels and replaced values deviating by more than 5$\sigma$ from the rescaled median spectrum by the corresponding median spectrum value, where $\sigma$ is the MAD scatter across integration. The resulting time-series spectra were used to construct the order~1 and order~2 spectroscopic light curves at the native detector-pixel resolution.

\subsection{Light-curve Fitting}
We fitted the spectroscopic light curves using the \texttt{juliet} package \citep{Espinoza_2019}, using \texttt{batman} transit models \citep{Kreidberg_2015} and \texttt{dynesty} nested-sampling algorithm \citep{Speagle_2020} for posterior distribution estimation. Orbital parameters were fixed to a joint analysis of JWST transit observations and radial-velocity measurements of the LHS\,1140 system (Salhi \& Gressier et al. in prep.) For LHS\,1140\,b, we adopted $P=24.73724423$\,days, $T_0=2460304.698258$~$\mathrm{BJD}_{\mathrm{TDB}}$, $a/R_\star=94.31$, $b=0.17$, $e=0.008144$, and $\omega=53.37^\circ$. For LHS\,1140\,c, we fixed  $P=3.777935859$\,days, $T_0=2460304.707760$~$\mathrm{BJD}_{\mathrm{TDB}}$, $a/R\star=26.95$, $b=0.19$, $e=0.003864$, and $\omega=-190.48^\circ$. We fixed $a/R_\star$ and $b$ for the single-planet visits. For the second DDT~6543 visit, which contains simultaneous transits of LHS\,1140\,b and c, we instead fixed the stellar density to $\rho_\star=25.9~\mathrm{g\,cm^{-3}}$ and modeled the two transit signals jointly, using independent planet-to-star radius ratios and common visit-dependent systematic parameters. Covariance between the two transit depths can contribute to the larger uncertainties on the recovered spectrum of LHS\,1140\,b. All remaining visits were modeled with a single transit signal. We adopted the prior $\mathcal{U}(0,0.2)$ for $R_{\rm p}/R_\star$.
%For the second DDT~6543 visit which contains simultaneous transits of LHS\,1140\,b and c; we instead fixed the stellar density to $\rho_\star=25.9~\mathrm{g\,cm^{-3}}$ and modeled the two transit signals jointly, using independent planet-to-star radius ratios and common visit-dependent systematic parameters. All remaining visits were modeled ith a single transit signal. We adopted the prior $\mathcal{U}(0,0.2)$ for $R_{\rm p}/R_\star$. 

For each light curve, we performed a \textit{conservative} and a \textit{fiducial} fit. In the \textit{conservative} fit, the quadratic limb-darkening coefficients were fitted using $u_1,u_2\sim\mathcal{U}(-3,3)$, while instrumental systematics and time-correlated noise were modeled with a Matérn-$3/2$ Gaussian process. We adopted log-uniform priors between ($10^{-8},10^{-2}$) for the GP amplitude and ($10^{-8},10^{2}$) for the timescale. The \textit{fiducial} fit used the native pixels from $1.00$ to $1.15,\upmu\mathrm{m}$ and fixed $u_1$ and $u_2$ to values inferred from the broadband light curve of the lowest-scatter visit, GO~7073 observation~21. The GP was replaced by a linear temporal slope with $\theta_0\sim\mathcal{U}(-10,10)$. The order~1 spectroscopic light curves were fitted at the native detector pixel resolution. The transit depth in each wavelength channel was calculated as $(R_{\rm p}/R_\star)^2$, with its value and uncertainty derived from the median and 16th--84th percentiles of the posterior.

\begin{table*}[htpb]
\centering
\caption{Helium constraints obtained with the \textit{fiducial} and \textit{conservative} light-curve treatments. The upper limits correspond to $3\sigma$, and the injection--recovery tests assume an intrinsic line with $A=1.24\%$ and $\mathrm{FWHM}=0.86$\,\AA.}
\label{tab:results}
\small
\renewcommand{\arraystretch}{1.08}
\setlength{\tabcolsep}{10pt}

\begin{tabular}{lcccc}
\hline
Parameter
& 2023-12-01
& 2023-12-25
& 2025-08-10
& 2026-07-23 \\
\hline

\multicolumn{5}{l}{\textbf{Fiducial light-curve treatment}} \\[2pt]

\multicolumn{5}{l}{\textit{SOSS-only low-resolution fit}} \\
$A_{\mathrm{LR}}$ [ppm]
& $-8.0^{+124.1}_{-125.3}$
& $+42.3^{+176.9}_{-177.1}$
& $+56.3^{+115.1}_{-114.3}$
& $+6.3^{+112.4}_{-110.4}$ \\

$Z_{0,\mathrm{LR}}$ [$\sigma$]
& $-0.06$
& $+0.24$
& $+0.49$
& $+0.06$ \\[2pt]

\multicolumn{5}{l}{\textit{WINERED-informed intrinsic-line fit}} \\
$A_{\mathrm{He}}$ [\%]
& $-0.01^{+0.26}_{-0.26}$
& $+0.09^{+0.37}_{-0.37}$
& $+0.12^{+0.24}_{-0.24}$
& $+0.02^{+0.23}_{-0.24}$ \\

Upper limit on $A_{\mathrm{He}}$ [\%]
& $0.79$
& $1.21$
& $0.84$
& $0.72$ \\

$Z_{0,\mathrm{He}}$ [$\sigma$]
& $-0.05$
& $+0.25$
& $+0.51$
& $+0.06$ \\

$Z_{\mathrm{W24-SOSS}}$ [$\sigma$]
& $3.59$
& $2.62$
& $3.39$
& $3.74$ \\[2pt]

%\multicolumn{5}{l}{\textit{Injection--recovery test}} \\
%$Z_{\mathrm{injection}}$ [$\sigma$]
%& $4.70$
%& $3.34$
%& $5.15$
%& $5.29$ \\

\hline

\multicolumn{5}{l}{\textbf{Conservative light-curve treatment}} \\[2pt]

\multicolumn{5}{l}{\textit{SOSS-only low-resolution fit}} \\
$A_{\mathrm{LR}}$ [ppm]
& $-160.2^{+219.2}_{-217.8}$
& $-15.1^{+180.8}_{-180.8}$
& $+55.6^{+134.6}_{-135.7}$
& $+63.2^{+131.5}_{-130.4}$ \\

$Z_{0,\mathrm{LR}}$ [$\sigma$]
& $-0.73$
& $-0.08$
& $+0.41$
& $+0.48$ \\[2pt]

\multicolumn{5}{l}{\textit{WINERED-informed intrinsic-line fit}} \\
$A_{\mathrm{He}}$ [\%]
& $-0.33^{+0.46}_{-0.45}$
& $-0.03^{+0.38}_{-0.38}$
& $+0.12^{+0.28}_{-0.28}$
& $+0.14^{+0.27}_{-0.28}$ \\

Upper limit on $A_{\mathrm{He}}$ [\%]
& $1.05$
& $1.11$
& $0.96$
& $0.96$ \\

$Z_{0,\mathrm{He}}$ [$\sigma$]
& $-0.73$
& $-0.08$
& $+0.42$
& $+0.51$ \\

$Z_{\mathrm{W24-SOSS}}$ [$\sigma$]
& $3.08$
& $2.87$
& $3.07$
& $3.08$ \\[2pt]

%\multicolumn{5}{l}{\textit{Injection--recovery test}} \\
%$Z_{\mathrm{injection}}$ [$\sigma$]
%& $2.71$
%& $3.24$
%& $4.34$
%& $4.45$ \\

\hline
\end{tabular}
\end{table*}

\section{Helium Analysis}\label{sec:helium}
From the spectroscopic light-curve fits, we extract the pixel-resolution transmission spectra 
from the four transits of LHS 1140\,b and two transits of LHS 1140\,c. We consider 21 to 22 pixels spanning $1.073$--$1.093\,\upmu\mathrm{m}$ which encompass the He~I triplet feature near $10\,833$\,\AA\ and provide sufficient baseline to constrain the local continuum.

\subsection{Low-resolution helium inference from SOSS}
We first searched for helium independently of the WINERED helium detection reported by \citetalias{Cherubim2026} in the September 2024 observations (hereafter W24). We modeled the He~I triplet as a single Gaussian centered at $10\,833.22$\,\AA, with an observed FWHM fixed to the SOSS resolution $\mathrm{FWHM}_{\mathrm{SOSS}}= \frac{\lambda_{\mathrm{He}}}{R}$, where $R=600$, similar to e.g., \cite{Fournier_Tondreau_2024,Fournier_Tondreau_2025, Ahrer2025}. The Gaussian profile was integrated over the wavelength boundaries of each SOSS bin. We searched for excess absorption in the transit depth using the following model in each bin $i$:
\begin{equation}
M_i = D_0 + A_{\mathrm{LR}} G_i,
\end{equation}
where $D_0$ is the local continuum in ppm, $A_{\mathrm{LR}}$ is the observed peak amplitude in ppm, and $G_i$ is the dimensionless bin-integrated SOSS resolution Gaussian profile normalized to a unit observed peak.

\subsection{WINERED-informed intrinsic helium inference}
We then tested how the W24 line profile would appear in the SOSS observations. We modeled the He~triplet as a single intrinsic Gaussian centered at $10833$\,\AA\ with a fixed full width at half-maximum (FWHM) of 0.86\,\AA\,, corresponding to the value reported in \citetalias{Cherubim2026}. This Gaussian represents the intrinsic line before convolution with the SOSS instrumental response. The line is actually unresolved with NIRISS/SOSS, so we convolved the intrinsic profile with a Gaussian line spread-function at $R=600$. The observed Gaussian width is then calculated as : $\sigma_{\mathrm{obs}}^2 = \sigma_{\mathrm{int}}^2 +\sigma_{\mathrm{inst}}^2$,
where $\sigma_{\mathrm{inst}}=(\lambda_{\mathrm{He}}/R)/2.355$. The amplitude of the signal is diluted but the integrated area is conserved. The convolved profile was integrated analytically over the wavelength boundaries of each SOSS bin. For each visit, we modeled the signal in each bin $i$ as :
\begin{equation}
M_i=D_0+\left(\frac{A_{\mathrm{He}}}{1\%}\right)T_i
\end{equation} where $D_0$ is a constant corresponding to the transit depth continuum in ppm, $A_{\mathrm{He}}$ is the amplitude of the He line depth in percent and $T_i$ is the bin-integrated SOSS signal in ppm produced by an intrinsic line depth of 1\%. \\

Both the low-resolution and the WINERED-informed models were fitted using the \texttt{emcee} sampler with 32 walkers, discarding the 2000 burn-in steps and retaining 10000 steps \citep{Foreman_Mackey_2019}. We adopted uniform priors of $\pm1000$\,ppm around the median transit depth for $D_0$, $\mathcal{U}(-3000,3000)\,$ppm for $A_{\mathrm{LR}}$ and  $\mathcal{U}(-3,3)\,\%$ for A$_{\mathrm{He}}$. Allowing negative amplitudes avoids a positive-boundary bias. We construct the posterior on each fitted quantity and report the median $A_{50}$, and the 16th--84th percentile interval. We first assessed the presence of helium independently in each observation by measuring the distance of the inferred amplitude from zero: 
\begin{equation}
Z_{0,X} = \frac{A_{\mathrm{X},50}}{(A_{\mathrm{X},84}-A_{\mathrm{X},16})/2}
\end{equation}

We then compared each SOSS constraint with the helium absorption reported by \citetalias{Cherubim2026}: $A_{\mathrm{W24}} =1.24^{+0.23}_{-0.22}\%$. For each SOSS observation, we calculated the difference between the posterior medians $\Delta A_{\mathrm{He}}$ = 1.24\% - $A_{50}$. Because all SOSS medians lie below the W24 value, we computed
\begin{equation}
Z_{\mathrm{W24-SOSS}}=\frac{A_{\mathrm{W24}}-A_{\mathrm{He},50}}{\sqrt{\sigma_{\mathrm{W24}-}^{2}+\sigma_{\mathrm{He}+}^{2}}},
\end{equation} 
where the asymmetric uncertainties extending toward the other measurement are combined in quadrature. The difference is then expressed in units of uncertainty to obtain a $\sigma$-away metric between the two measurements. Values above $3$ indicate that W24-like absorption is disfavored at more than $3\sigma$, and thus that an excess absorption similar to the one published in \citetalias{Cherubim2026} is disfavored in the individual SOSS observation. \\ 

As a validation we performed a spectral-level injection-recovery test using the wavelength grid and uncertainties of each visit. We generated spectra containing an intrinsic Gaussian with $A_{\rm inj}=1.24\%$ and $\mathrm{FWHM}=0.86$ \AA\, convolved and binned to the SOSS resolution. We analyzed the spectra with the same inference procedure. These tests confirm that such a profile remains detectable after convolution and binning.

\begin{figure*}[htpb]
    \centering
        \includegraphics[width=\textwidth]{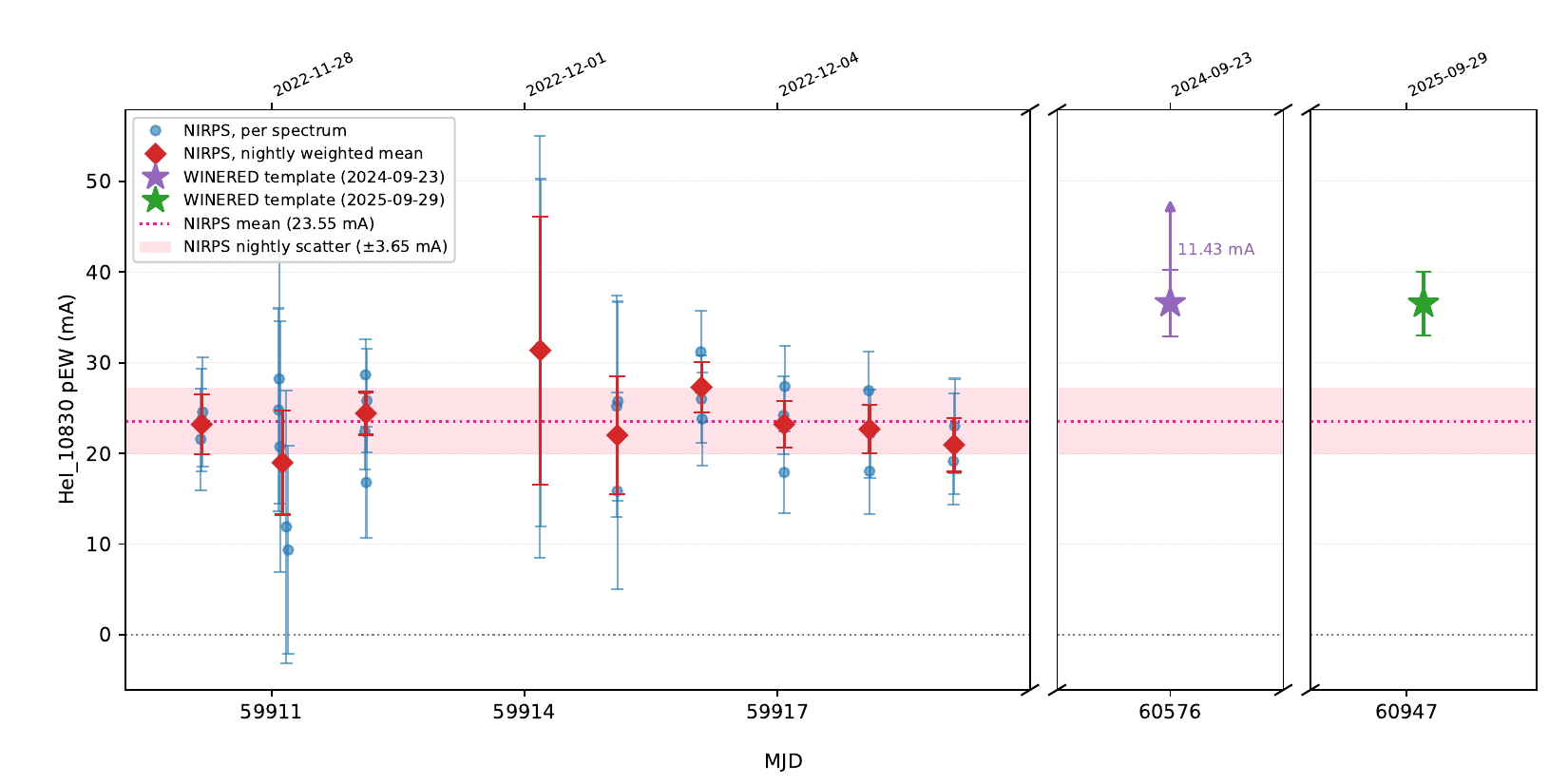}
    \caption{He\,I 10830-\AA\ EW time series for LHS\,1140. Blue circles show 29 individual NIRPS spectra from nine nights in 2022, and red diamonds their nightly weighted means. The purple and green star marks the \citetalias{Cherubim2026} template measurement on 2024 September 23 and 2025 September 29th respectively. The  arrow of the  2024 event indicates the excess absorption observed during that visit. The full sequence spans a range in He\,I EW from approximately 20 to 50\,m\AA.}
    \label{fig:heI_timeserie}
\end{figure*}

\section{Results and Discussion}\label{sec:results}
\subsection{SOSS constraints on helium absorption}
We analyzed the four NIRISS SOSS observations of LHS\,1140\,b independently for each of the two light-curve treatments (see Table~\ref{tab:results}). Neither the low-resolution search nor the WINERED-informed intrinsic-line fit provides evidence for positive helium absorption. For the \textit{fiducial} treatment the respective amplitude deviates from zero at most by $+0.49\sigma$ for the low-resolution fit and by $+0.51\sigma$ for the WINERED-informed fit, while the GP treatment yields similar results. Figure~\ref{fig:helium_intrinsic_posteriors} compares the four continuum-subtracted SOSS transmission spectra with the WINERED-informed helium model and shows the intrinsic amplitude posteriors. The inferred helium amplitudes range from $-0.01\%$ to $+0.12\%$, with $3\sigma$ upper limits of $0.72$ to $1.21\%$. The individual SOSS constraints differ from the intrinsic helium depth reported by \citetalias{Cherubim2026} by $2.62$--$3.74\sigma$, with three visits exceeding $3\sigma$. The least precise constraint is obtained for the 2023 December 25 cotransit of planets b and c. From the injection test, the probability of recovering the W24 signal above $3\sigma$ ranges from $63.8\%$ for the 2023 December 25 observation to more than $95\%$ for each of the other three visits. 

Assuming that the intrinsic helium depth remained constant across the four SOSS epochs, we multiplied the independent visit posterior. The resulting joint constraint is $A_{\mathrm{joint}}=0.05\pm0.13\%$, consistent with zero within $0.4\sigma$ (see Figure~\ref{fig:helium_intrinsic_posteriors} (right)). A signal as strong as that reported by \citetalias{Cherubim2026} is disfavored at $4.5\sigma$. 

We also applied both analyses to the two SOSS transits of LHS\,1140\,c and found no evidence for helium absorption. For the fiducial treatment, the low-resolution amplitudes are $A_{\mathrm{LR}}=-161^{+212}_{-209}$\,ppm and $-132^{+137}_{-137}$\,ppm for the DDT~6543 and GO~7073 visits, respectively. The corresponding WINERED-informed amplitudes are $A_{\mathrm{He}}=-0.33^{+0.44}_{-0.44}\%$ and $-0.28^{+0.28}_{-0.28}\%$, with $3\sigma$ upper limits of $1.02\%$ and $0.58\%$. All measurements are consistent with zero within $1\sigma$.

We also inspected the one- and two-pixel spectroscopic light curves centered on the He\,I triplet. None of the visits shows coherent flux decrease during the extended pretransit (0.84hr) or posttransit (0.75hr) intervals reported by \citetalias{Cherubim2026}. We therefore find no evidence for temporally extended He\,I absorption in the NIRISS/SOSS observations.

\subsection{Search for Stellar flare with H-alpha}
We searched for contemporaneous stellar flares in four SOSS observations, including the visit of LHS\,1140\,c alone, using H$\alpha$ in order\,2. We found no enhancement that was both temporal and localized at the H$\alpha$ line. There is no convincing evidence for flares in any visits. Details of the methods and results are presented in Appendix~\ref{app:halpha} and Figure~\ref{fig:Halpha_spectra}. Because order\,2 has a lower SNR and is affected by field-star contamination in some observations, this diagnostic should nevertheless be interpreted cautiously.

\subsection{Occurrence rate of helium}
Considering that one event at the level measured with WINERED was detected out of six transit observations, we modeled the occurrence of such events as a binomial process. The probability of observing $k$ events among $N$ independent transits is 
\begin{equation}
P(k \mid f, N) = \binom{N}{k} f^{k}(1-f)^{N-k}
\end{equation}
where $f$ is the probability of observing a WINERED 2024-like %Cherubim-like 
helium absorption event during a transit. For $k=1$ and $N=6$, and a uniform prior [0,1] on f, we find $f = 22_{-12}^{+17}\%$. Therefore, if the high resolution detection is indeed planetary, the rate of such atmospheric loss events must be relatively low. 

\subsection{Helium variability in LHS 1140 }
\label{sect:heliumvar}
%nondetection of the helium triplet in transmission spectroscopy does not rule out the presence of helium in a planetary atmosphere, 
The SOSS nondetections show that helium absorption at the level measured in September 2024 is not persistent, but do not by themselves determine its origin. A nondetection of metastable He\,I does not exclude a helium-bearing atmosphere but rather that the metastable state is not populated. This state highly depends on the XUV to NUV flux received by the planet \citep{Seager2000,oklopcic2018, oklopcic2019}. Therefore, one possible interpretation for these nondetections is that the 2024 September detection with WINERED was associated with a transient increase in the X-ray and extreme-ultraviolet (XUV) irradiation of the planet, which could temporarily enhance the metastable helium population and atmospheric escape.
Alternatively, temporal variations in the stellar He\,I line could contribute to the excess measured in 2024. 
%\citep[e.g.,][]{Sanz-Forcada2025}. 
%An alternative possibility is that the signal measured with WINERED originated from variability of the stellar He\,I line itself. %This possibility is particularly important because NIRISS/SOSS has sufficient sensitivity to detect a change in the integrated He\,I absorption of the magnitude reported by \citetalias{Cherubim2026}, but insufficient spectral resolution to resolve the stellar He\,I line to characterize its epoch-to-epoch variability.

To explore the latter possibility, we reanalyzed the out-of-transit WINERED stellar spectra obtained on 2024 September 23 and 2025 September 29 and made available by \citetalias{Cherubim2026}. Details are provided in the Appendix~\ref{app:nirps}. We measure consistent He\,I equivalent widths of $36.6\pm4.8$ and $36.5\pm4.6$\,m\AA\  in 2024 and 2025 respectively. For the 2024 spectrum, we obtained a stellar-line FWHM of $0.51\pm0.11$\,\AA\ , only $1.2\sigma$ narrower than the GP-based in-transit FWHM of $0.86^{+0.15}_{-0.27}$\,\AA\  reported by \citetalias{Cherubim2026}. The Supplementary Materials of \citetalias{Cherubim2026} instead report a $\sim$9$\sigma$ difference based on fits without a GP, for which the formal width uncertainties are substantially smaller.\footnote{The line-width values labeled as FWHM in the Supplementary Materials of \citetalias{Cherubim2026} are in fact Gaussian $\sigma$ and should therefore be multiplied by $2.355$ to obtain the corresponding FWHM, following clarification from C.\ Cherubim.}

%To investigate this possibility, we examined the stellar He\,I absorption of LHS\,1140 at multiple epochs. We first reanalyzed the out-of-transit WINERED stellar spectrum obtained on 2024 September 23 and 2025 September 29 2025 by \citetalias{Cherubim2026} available through Zenodo. Details of this analysis are presented in the Appendix. We find consistent out-of-transit He\,I equivalent widths in 2024 and 2025, with values of $36.6\pm4.8$ and $36.5\pm4.6$~m\AA, respectively. For the 2024 spectrum, we measure a FWHM of $0.51\pm0.11$~\AA, only marginally narrower ($1.2\sigma$) than the in-transit FWHM of $0.86^{+0.15}_{-0.27}$~\AA\ reported by \citetalias{Cherubim2026}. This marginal difference is at odds with the much higher (9$\sigma$) significance reported by \citetalias{Cherubim2026} from a fit performed without a Gaussian-process treatment of the correlated noise; notably, the corresponding no-GP in-transit width uncertainty is nearly an order of magnitude smaller than in their GP-based analysis.\footnote{According to information provided by first author C.\ Cherubim, the line-width values reported as FWHM in the  Supplementary Materials of \citetalias{Cherubim2026} are in fact Gaussian $\sigma$ values, and should therefore be multiplied by $2.355$ to obtain the corresponding FWHM.}

We also analyzed 29 archival NIRPS \citep{Bouchy2025} spectra obtained over nine commissioning nights in late 2022, approximately two years before the WINERED observations. As shown in Figure~\ref{fig:heI_timeserie}, the NIRPS spectra yield a mean stellar He\,I EW of $23.7$~m\AA\  with a night-to-night dispersion of $3.5$~m\AA, slightly larger than median nightly error bars of $2.97$~m\AA\.  The out-of-transit WINERED measurements of $\sim$37\,m\AA\ in both 2024 and 2025, are $2.2\sigma$ above the NIRPS measurement providing tentative evidence for epoch-to-epoch variability. %are above the exceeds the NIRPS measurement by $2.2\sigma$. 
Adding the W24 excess absorption of $11.3^{+2.8}_{-3.7}$~m\AA\ to the contemporaneous stellar line gives an in-transit EW of $47.9^{+5.6}_{-6.3}$~m\AA, 3.4-$\sigma$ higher than the NIRPS's average signal.%Including the planetary excess absorption of $11.3^{+2.8}_{-3.7}$~m\AA\ measured by \citetalias{Cherubim2026} in 2024, the inferred in-transit EW is $47.9^{+5.6}_{-6.3}$~m\AA, 3.4-$\sigma$ higher than the NIRPS's average signal. 
 Thus, the available observations of LHS\,1140 span approximately $20$--$50$~m\,\AA\ in He\,I EW. 

Such variability is not unexpected for a star as cool as LHS\,1140. We compared these measurements with the CARMENES M-dwarf samples of \citet{Fuhrmeister2020} restricting the comparison to valid measurements and excluding young stars (Appendix~\ref{app:carmenes_hei}). For each CARMENES target, the mean pseudo He\,I EW and its variability were characterized from multiepoch observations, typically comprising a few tens of spectra per star. As noted by \citet{Fuhrmeister2020}, the strength of the He\,I line decreases systematically toward later spectral types and both the NIRPS and WINERED measurements of LHS\,1140 lie within the general downward trend toward lower $T_{\rm eff}$ 
(See Appendix\,\ref{app:carmenes_hei}; Figure~\ref{fig:combined}, left panel). More importantly, the fractional variability, $\sigma_{\rm pEW}/\mathrm{pEW}$, increases significantly toward lower $T_{\rm eff}$ (Fig.~\ref{fig:combined}, right panel) with  LHS\,1140 from the NIRPS and WINERED observations following this same trend. Thus, both the mean He\,I strength and the level of variability observed in LHS\,1140 are consistent with the empirical trend established from early to mid M dwarfs monitored over multiple epochs. Our multiepoch analysis reveals for the first time that a mid-to-late M dwarf like LHS\,1140 can show significant variability in the stellar He\,I line.

One caveat with the above analysis comparing measurements obtained with different instruments is the possibility of systematic differences arising from uncertainties in defining the pseudo-continuum and from differences in telluric correction procedures. To minimize the former, we analyzed both the NIRPS and WINERED spectra using the same methodology as \citetalias{Cherubim2026}, fitting a polynomial to the pseudo-continuum over the same wavelength interval (gray-shaded region in Fig.~\ref{fig:HeI_NIRPS_WINRED_template}). As an additional check, we measured the time series of four other absorption lines within the WINERED spectral range, at 1082.79, 1082.93, 1083.18, and 1083.23 nm, as described in the Appendix~\ref{app:template} (see Fig.~\ref{fig:HeI_NIRPS_WINRED_template_other_lines}). These measurements reveal no evidence of a systematic bias that would cause the He\,I equivalent widths measured with NIRPS to be systematically smaller than those obtained with WINERED. Nevertheless, we caution that future investigations of the long-term variability of the He\,I signal in LHS\,1140 would benefit from repeated observations with the same instrument to minimize potential inter-instrument systematics.

%As described in more details in the Appendix, Figure\,\ref{fig:combined} compares its He\,I EW with the CARMENES M-dwarf sample of \citet{Fuhrmeister2020}, comprising only "valid" measurements and excluding young stars. For each CARMENES target, the mean pseudo He\,I EW and its variability were characterized from multiepoch observations, typically comprising a few tens of spectra per star. %\footnote{The ``pseudo'' EW is measured relative to a locally defined pseudo-continuum rather than a true continuum, which is difficult to identify in the heavily line-blanketed spectra of M dwarfs.}
%As noted by \citet{Fuhrmeister2020}, the strength of the He\,I line decreases systematically toward later spectral types and both the NIRPS and WINERED measurements of LHS\,1140 lie within the general downward trend toward lower $T_{\rm eff}$ 
%(See Appendix~\ref{app:carmenes_hei}; Figure~\ref{fig:HeI_left}). More importantly, the fractional variability, $\sigma_{\rm pEW}/\mathrm{pEW}$, increases significantly toward lower $T_{\rm eff}$ (Fig.~\ref{fig:HeI_right}) with  LHS\,1140 from the NIRPS and WINERED observations following this same trend. Thus, both the mean He\,I strength and the level of variability observed in LHS\,1140 are consistent with the empirical trend established from early to mid M dwarfs monitored over multiple epochs. Our multiepoch analysis reveals for the first time that a mid-to-late M dwarf like LHS\,1140 can show significant variability in the stellar He\,I line.

For transmission spectroscopy, the relevant question is whether stellar He\,I variability can occur on timescales comparable to a transit (over several hours). High-cadence Sun-as-a-star NIRPS observations reveal variations of the He\,I triplet on timescales from minutes to days, including changes on several-hour timescales comparable to the duration of a transit (see Figure~3 of \citealt{Mercier2025}).
%obtained with NIRPS show that this is occurring for the Sun. Indeed, \citet{Mercier2025} found significant variability of the solar He\,I triplet over timescales ranging from minutes to days, including changes on several-hour timescales comparable to the duration of a transit observation (see Figure~\ref{fig:heI_timeserie} of \citealt{Mercier2025}).
Over two multi-hour sequences, they find that the solar He I EW varies by $\sim$10\% peak-to-peak, with correlated structure that can mimic a transit-like signal. As discussed above, the W24 transit shows a fractional increase of 30\% in the He I EW. Extrapolating the empirical relation for the CARMENES M-dwarf sample to cooler stars suggests that variability of this amplitude may be possible (Fig.~\ref{fig:combined}(right)). However, it remains unknown whether fully convective M dwarfs can exhibit such variability on short timescales, and therefore whether the W24 signal could be attributed to stellar variability.
\begin{figure*}[htpb]
    \centering
        \includegraphics[width=\textwidth]{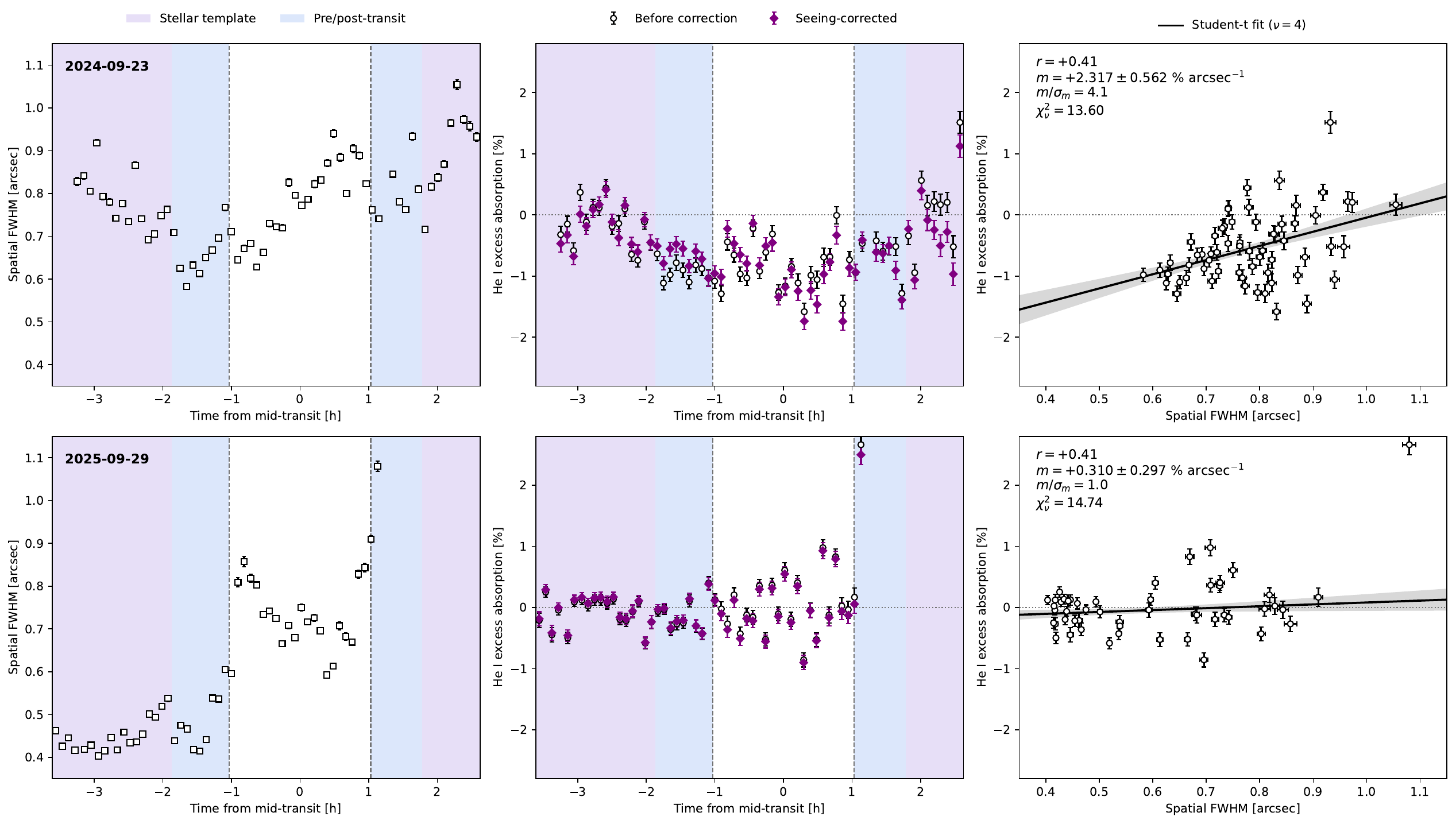}
    \caption{Potential dependence of the 2024 and 2025 WINERED He\,I signal on observing conditions, (top and bottom panels respectively).
    \textit{Left panels:} Image quality (seeing) during the observing sequence; shaded intervals follow \citetalias{Cherubim2026}, with pre- and posttransit measurements in blue and measurements used to construct the stellar OOT template in purple. The dashed lines represents the transit window. \textit{Middle panels:} He\,I excess absorption as shown in Figure~3A of \citetalias{Cherubim2026} (white) and the corresponding seeing-corrected measurements (dark purple). 
     \textit{Right panels:} He\,I excess absorption versus measured seeing FWHM for each exposure. The solid line shows the robust linear fit described in the text, revealing a moderate positive correlation with seeing in 2024 ($4.1\sigma$ significance) but marginal ($1.0\sigma$) in 2025. The relatively large reduced $\chi^2$ is indicative of under-estimated error bars.}
    \label{fig:heI_vs_seeing}
\end{figure*}

\subsection{Potential instrumental systematics contribution to the W24 He I Signal}
Beyond the planetary and stellar interpretations considered above, we also explored whether instrumental systematics could have contributed to the W24 signal. The line-profile analysis of \citetalias{Cherubim2026} adopts a fixed instrumental resolution throughout the sequence. Seeing-dependent changes in the illumination of the 0.3 arcsec-wide WINERED slit \citep{Bessho2022} could therefore modify the effective spectral resolution and, in turn, the measured depth of the narrow He\,I feature.  

The seeing was estimated for each measurement in the 2024 and 2025 sequences available through the accompanying Zenodo repository by fitting a Gaussian to the spatial profile of the spectral trace and converting to arcseconds using the WINERED plate scale of $0.15''$/pixel \citep{Bessho2022}.
As shown in Figure~\ref{fig:heI_vs_seeing}, the spatial FWHM, used here as a proxy for seeing, was overall better in 2025 but varied more significantly by a factor of $\sim$2, compared to $\sim$20\% in 2024. 
The He\,\textsc{I} excess absorption was fitted against the FWHM by maximum likelihood, as a straight line plus a constant excess scatter $s$. Each frame follows a Student-$t$ distribution ($\nu = 4$) with a squared scale of
$s^2 + \sigma_i^2$, where $\sigma_i$ is its He\,\textsc{I} error, so that outliers
are down-weighted rather than rejected. The slope error is taken from the covariance matrix at the likelihood maximum, and is propagated into the error on the corrected He\,\textsc{I} excess together with the FWHM errors. This method is thus resilient to outliers. The He\,I absorption excess shows a moderate positive correlation with seeing ($r=+0.41$; right panels of Figure~\ref{fig:heI_vs_seeing}), with slopes differing from zero at $4.1\sigma$ and $1.0\sigma$ for the 2024 and 2025 data, respectively. The large reduced chi-square values ($\chi_\nu^2=13$--$14$) in both sequences suggest substantial excess variability, potentially due to correlated noise in the He\,I absorption excess as mentioned in \citetalias{Cherubim2026}. For the W24 sequence, correcting each measurement to the mean spatial FWHM reduces the He\,I excess over the combined pre-, in-, and posttransit interval relative to the OOT baseline by approximately $20\%$. The empirical correlation between the He\,I signal and seeing does not establish an instrumental origin for the W24 signal, as it may be coincidental given their similar, approximately parabolic temporal trends. The correlation is much weaker in W25. More broadly, the excess variability in both sequences compared to individual measurement uncertainties, potentially arising from correlated noise, warrants further investigation.

Quantifying this contribution would require a detailed characterization of the time-dependent spectral resolution in both WINERED sequences and is beyond the scope of this work. Measuring the signal through its equivalent width would provide a useful complementary test because equivalent width is less sensitive than line depth to changes in spectral resolution. The middle top panel of Figure~\ref{fig:heI_vs_seeing}, adapted from Figure~3A of \citetalias{Cherubim2026}, also illustrates the sensitivity of the inferred contrast to the adopted OOT baseline. The pre- and posttransit intervals were interpreted as extended planetary absorption and therefore are excluded from the stellar template. If part of this variability instead has a stellar or instrumental origin, including these intervals in the baseline would reduce the apparent transit contrast.

Taken together, the detection in only one of the two WINERED observations and the four SOSS nondetection show that a W24-like absorption is not persistent. The epoch-to-epoch variability of the stellar He\,I line and the lack of confirmation of significant in-transit line broadening indicate that a stellar contribution remains possible, consistent with the behavior of early- to mid-M dwarfs in the CARMENES sample, and solar high-cadence observations of He\,I. The seeing variation during the 2024 WINERED sequence further motivate consideration of a possible instrumental contribution. These diagnostics do not rule out intermittent planetary escape, but indicate that stellar variability and instrumental systematics cannot presently be excluded as contributors the W24 signal. Further analysis of the WINERED data and new high-resolution observations, both in and out of transit, will be required to establish the planetary origin of the He\,I absorption in LHS~1140\,b.

\section{Summary and Conclusions}\label{sec:conclusions}

We searched for metastable He\,I absorption in four JWST/NIRISS SOSS transits of LHS\,1140\,b and found no evidence for excess absorption in any visit. Combining the four epochs, a persistent helium signal at the level reported by \citetalias{Cherubim2026} is disfavored at $4.5\sigma$. If the 2024 WINERED detection is planetary, the occurrence rate of such strong escape events must therefore be relatively low, $f=22^{+17}_{-12}\%$. 

The available near-infrared high-resolution spectra also show that the stellar He\,I line of LHS\,1140 is variable. Both its equivalent width and fractional variability are consistent with the empirical trend observed in early to mid-M dwarfs. Moreover, high-cadence NIRPS observations of the Sun demonstrate that He\,I variability can occur on several-hour timescales comparable to a transit observation of LHS\,1140\,b. The 2024 WINERED observations may have been affected by seeing-dependent variations in spectral resolution, potentially biasing the reported He\,I signal. An independent search for contemporaneous flaring using H$\alpha$ in SOSS order\,2 observations reveals no convincing activity signature. 

Our results therefore leave three viable interpretations of the 2024 WINERED signal: intermittent atmospheric escape from LHS\,1140\,b, a contribution from intrinsic stellar He\,I variability or instrumental effects. New high-resolution observations spanning both in- and out-of-transit phases and more in-depth analysis of the WINERED data are required to distinguish between these scenarios.

%\vspace{10mm}
%\nolinenumbers
%\begin{acknowledgments}
\section*{Acknowledgments}
R.D. would like to thank Collin Cherubim for insightful discussion on LHS~1140\,b and  his helpful guidance to use the WINERED data from the Zenodo repository (\href{https://doi.org/10.5281/zenodo.15723778}{zenodo.15723778}). 
This work is based on observations made with the NASA/ESA/CSA James Webb Space Telescope. The data were obtained from the Mikulski Archive for Space Telescopes (MAST) at the Space Telescope Science Institute (STScI) and are associated with programs 6543 and 7073. STScI is operated by the Association of Universities for Research in Astronomy, Inc., under NASA contract NAS 5-03127. The JWST NIRISS SOSS observations analyzed in this work are available from MAST at \dataset[10.17909/chgv-dj96]{https://doi.org/10.17909/chgv-dj96}
A.G. acknowledges support from the Trottier Family Foundation through the Trottier Postdoctoral Fellowship at the Institute for Research on Exoplanets (IREx).
C.C. acknowledges the support from the Swiss National Science Foundation under the grant SPECTRE (No 200021\_215200)
L.-P.C. acknowledges financial support from Mitacs through the Mitacs Accelerate program, in partnership with the Montreal Planetarium. R.D. and F.B. acknowledge financial support for the NIRPS project from the Canada Foundation for Innovation and the Swiss National Science Foundation through the PlanetS project that made the NIRPS project a reality. 
R.A. acknowledges the Swiss National Science Foundation (SNSF) support under the Post-Doc Mobility grant P500PT\_222212 and the support of IREx.
R.D.\ acknowledges financial support from the Natural Science and Engineering and Research Council of Canada, Canada Research Chair, the Trottier Family Foundation through IREx and Observatoire du Mont-M\'egantic.  
E.A. acknowledges support from the Trottier Family Foundation through IREx and Observatoire du Mont-M\'egantic.

\bigskip 
\noindent
\textit{Software:} \texttt{Astropy} \citep{astropy:2013,astropy:2018,astropy:2022}, \texttt{batman} \citep{Kreidberg_2015}, \texttt{emcee} \citep{Foreman_Mackey_2019}, \texttt{Matplotlib} \citep{Hunter_2007}, \texttt{NumPy} \citep{Harris_2020}, and \texttt{SciPy} \citep{2020SciPy-NMeth}.

%\clearpage

\begin{appendix}
\section{Search for H-alpha in SOSS observations} \label{app:halpha}
\begin{figure*}[htpb]
    \centering
        \includegraphics[width=14cm]{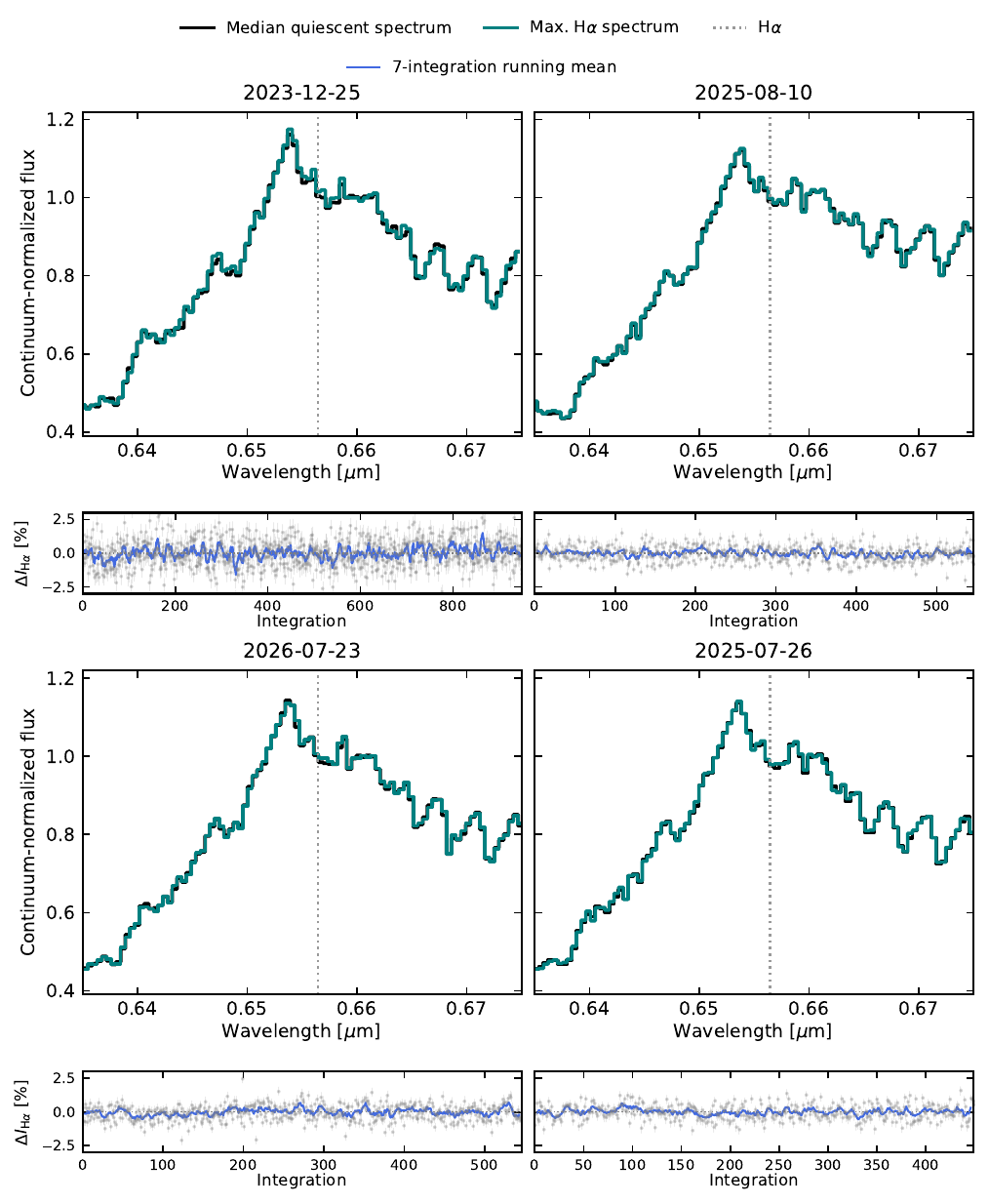}
        \caption{Comparison of the H$\alpha$ region in four JWST/NIRISS SOSS observations. Each main panel shows the median quiescent order-2 spectrum (black) and the median spectrum over seven integrations centered on the maximum of the H$\alpha$ luminosity (teal). Each integration was normalized using the adjacent continuum bands. The panel below shows the relative H$\alpha$-index excess time series (gray) and its seven-integration running mean (blue). }
    \label{fig:Halpha_spectra}
\end{figure*}

To investigate variability in the stellar H$\alpha$ line, we analyzed four JWST/NIRISS SOSS observations of the LHS\,1140 system. Three were obtained during transits of LHS\,1140\,b, on 2023 December 25, 2025 August 10, and 2026 July 23, while the fourth targeted a transit of LHS\,1140\,c on 2025 July 26. The first DDT~6543 observation was excluded because the displaced spectral traces and field-star contamination prevent a reliable analysis of order~2. For each remaining visit, we normalized every integration using continuum bands adjacent to H$\alpha$ and calculated an index from the ratio of the line flux to the continuum flux. The quiescent spectrum was constructed as the median of integrations with indices within three robust standard deviations of the visit median. We identified the maximum H$\alpha$ excess from the index time series and constructed the corresponding spectrum from the median of seven integrations centered on this maximum. We then compared the quiescent and maximum-excess spectra and inspected the index as a function of time. Figure~\ref{fig:Halpha_spectra} compares the quiescent and maximum-excess H$\alpha$ spectra from four JWST/NIRISS SOSS observations of the LHS\,1140 system. None of these visits shows a clear enhancement localized to the H$\alpha$ line and associated with a distinct event in the time series. %None of the visits shows an enhancement that is both temporally and localized to the H$\alpha$ line. 

\providecommand{\ion}[2]{#1\,\textsc{#2}}

\section{NIRPS data reduction and He I equivalent-width measurements}
\label{app:nirps}

\subsection{APERO reduction}
\label{app:apero}

The archival spectra of LHS\,1140 used in Section~\ref{sect:heliumvar} consist of 29 exposures of 600\,s obtained with NIRPS \citep{Bouchy2025} in high-efficiency mode ($R\approx75\,000$) over nine nights between 2022 November 27 and 2022 December 6, during the instrument commissioning. NIRPS is the near-infrared ($0.98-1.8\,\mu$m) high-resolution spectrograph installed alongside HARPS on the ESO 3.6m telescope at La Silla.

We used the products of version 0.7.292 of \texttt{APERO} \citep{Cook2022}, the pipeline originally developed for SPIRou and adapted to NIRPS. Briefly, \texttt{APERO} applies detector-level corrections to the raw images (bad pixels, background, non-linearity; \citealt{Artigau2018}), localizes the echelle orders, corrects the geometric distortions of the image plane, applies the flat-field and blaze corrections, and derives a wavelength solution from a combination of a hollow-cathode UNe lamp and a Fabry-P\'erot etalon \citep{Hobson2021}, using a mix of nightly and reference calibrations. Optimal extraction then produces two-dimensional spectra of 75 orders by 4088 pixels. The \ion{He}{1} triplet falls in order 15 (1073.9$-$1089.0\,nm).

\texttt{APERO} also handles both telluric absorption, through a fitted atmospheric transmission model \citep{Bertaux2014} followed by an empirical residual correction, and sky emission, dominated near 1083\,nm by OH lines, which is modeled from dedicated sky observations and subtracted. All spectra analyzed here are telluric- and sky-corrected, and divided by the blaze.

\subsection{Median template} \label{app:template} 
For each target, \texttt{APERO} also builds a template from all of its telluric-corrected spectra: the individual spectra are shifted onto a common rest wavelength grid (correcting for the barycentric Earth radial velocity), normalized order by order, and combined with a median along the time axis. We used the one-dimensional, constant-velocity-step version of this product, which for LHS\,1140 combines the 29 spectra listed above, as our high signal-to-noise reference spectrum of the star; it is the spectrum shown in the top panel of the figure of Sect.~\ref{app:ew}.

\subsection{Equivalent width of the He I line} \label{app:ew}
Equivalent widths were measured following the approach adopted for the CARMENES M-dwarf sample by \citet{Fuhrmeister2019, Fuhrmeister2020}: a linear pseudo-continuum and a small number of line components are fit jointly over the 1082.9$-$1083.5\,nm (10829$-$10835\,\AA) window, so that no separate continuum band is required. Spectra are first shifted to the stellar rest frame using the barycentric velocity and the systemic velocity of the target. The model is
\begin{equation}
M(\lambda) = \left(c_0 + c_1\lambda\right)
\left[1 + \sum_{k} A_k \, e^{-(\lambda-\lambda_k)^2 / 2\sigma_k^2}\right],
\end{equation}
with three components: the two unidentified absorption features immediately blueward of helium, and the \ion{He}{1} line itself, that is, the blend of the two reddest, unresolved components of the triplet (10833.22 and 10833.31\,\AA). Two differences with respect to \citet{Fuhrmeister2019} are worth noting. We do not include their photospheric \ion{Si}{1} component, whose width never converged on our data, and we use Gaussian rather than Voigt profiles, as the Lorentzian wings were systematically pushed against their bounds and were not required by the data.

The fit is performed in two stages. Individual 600\,s exposures do not have the signal-to-noise ratio needed to constrain the position and width of each component, so we first fit the median template of the star with all eleven parameters free (two continuum coefficients plus an amplitude, a center, and a width per component). The centers and widths derived from the template are then held fixed for every individual spectrum, whose fit therefore has only five free parameters (two continuum coefficients and three amplitudes), so that only the line amplitudes vary from epoch to epoch. For LHS\,1140, the template fit gives a \ion{He}{1} component with a depth of $5.37\pm0.65$\% and a FWHM of $0.0414\pm0.0058$\,nm.

The pseudo-equivalent width of the helium component alone, deblended from its two neighbors, is then obtained analytically from its Gaussian parameters. Its uncertainty follows the prescription of \citet[Sect. 3.3]{Fuhrmeister2019}: the amplitude of the helium component is fixed
at a grid of trial values around the best fit, the remaining parameters are refit at each trial value, a parabola is fit to the resulting $\chi^2(A)$ curve, and the uncertainty is taken as the amplitude offset for which $\chi^2$ increases by one. The nightly values shown in Figure~\ref{fig:heI_timeserie} are inverse-variance weighted means of the individual measurements.

\begin{figure*}[htpb]
    \centering
\includegraphics[
    height=0.8\textheight,
    keepaspectratio
]{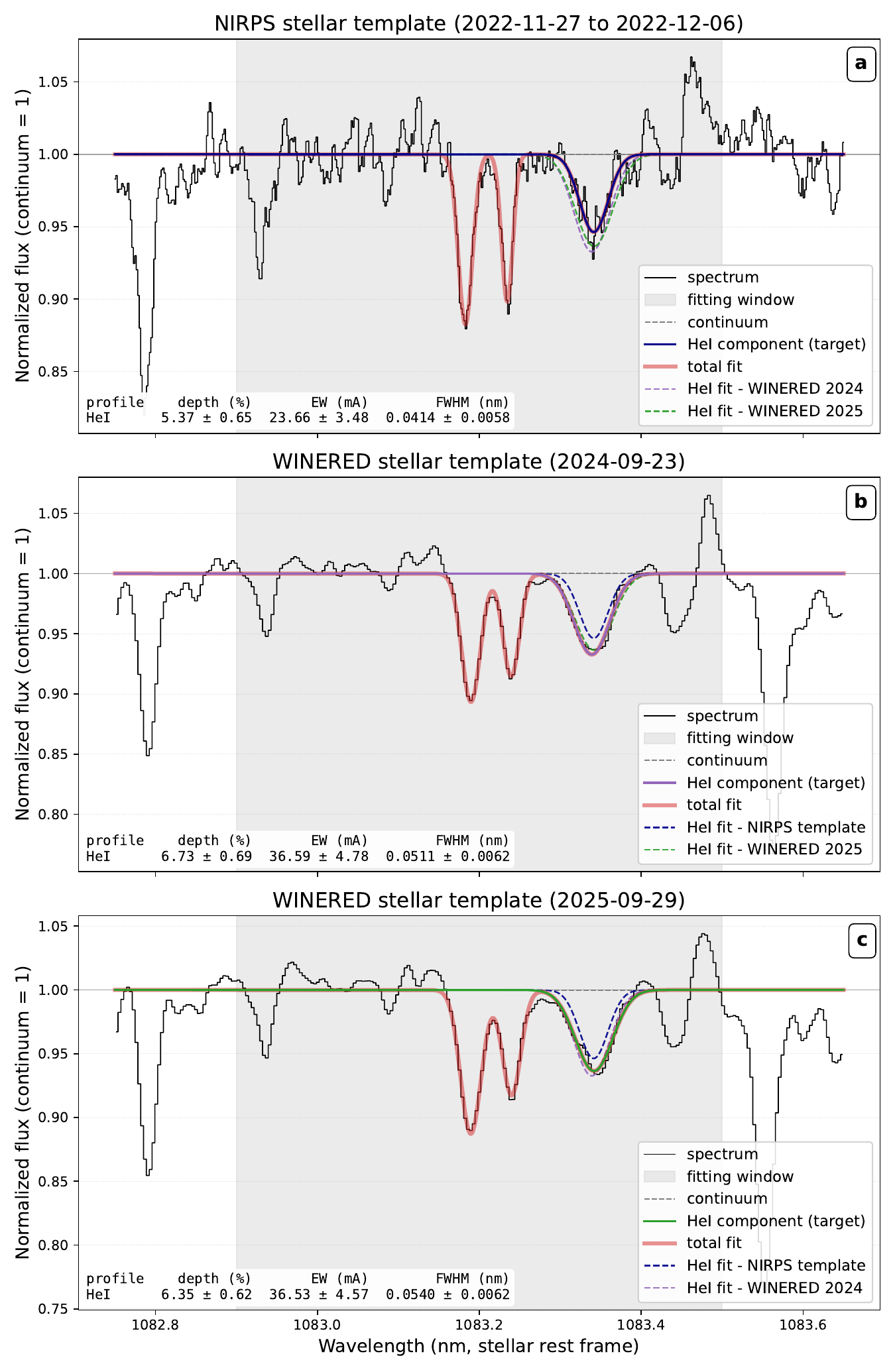}
    \caption{
    Comparison of the stellar He\,I 10830-\AA\ absorption profile of LHS\,1140 at three epochs; fitted parameters are given in the legend of each panel, and the gray shaded region indicates the fitting window.    
    Panel (a) shows the NIRPS stellar template from 2022 November--December, while panels (b) and (c) show the WINERED templates obtained during the 2024 September 23 and 2025 September 29 transits, respectively. 
    The 2024 and 2025 stellar He\,I profiles are mutually consistent despite excess planetary He\,I absorption being detected only in 2024, but both are significantly stronger than the NIRPS profile measured in 2022.
    }
    \label{fig:HeI_NIRPS_WINRED_template}
\end{figure*}

The same window, the same three components and the same error prescription were applied to the out-of-transit WINERED template of \citetalias{Cherubim2026} (Fig.~\ref{fig:HeI_NIRPS_WINRED_template}, bottom panel), the only difference being that the centers and widths were re-derived on that spectrum, since it comes from a different instrument. That fit gives a depth of $6.73\pm0.69$\% and a FWHM of $0.0511\pm0.0062$\,nm. All fits use \texttt{scipy.optimize.curve\_fit} \citep{Virtanen2020} in its bounded form, with physically motivated bounds on the amplitudes, widths, and center offsets; the parabolic fit to $\chi^2(A)$ uses \texttt{numpy.polyfit} \citep{Harris2020}.

\begin{figure*}[htpb]
    \centering
\includegraphics[
    height=0.8\textheight,
    keepaspectratio
]{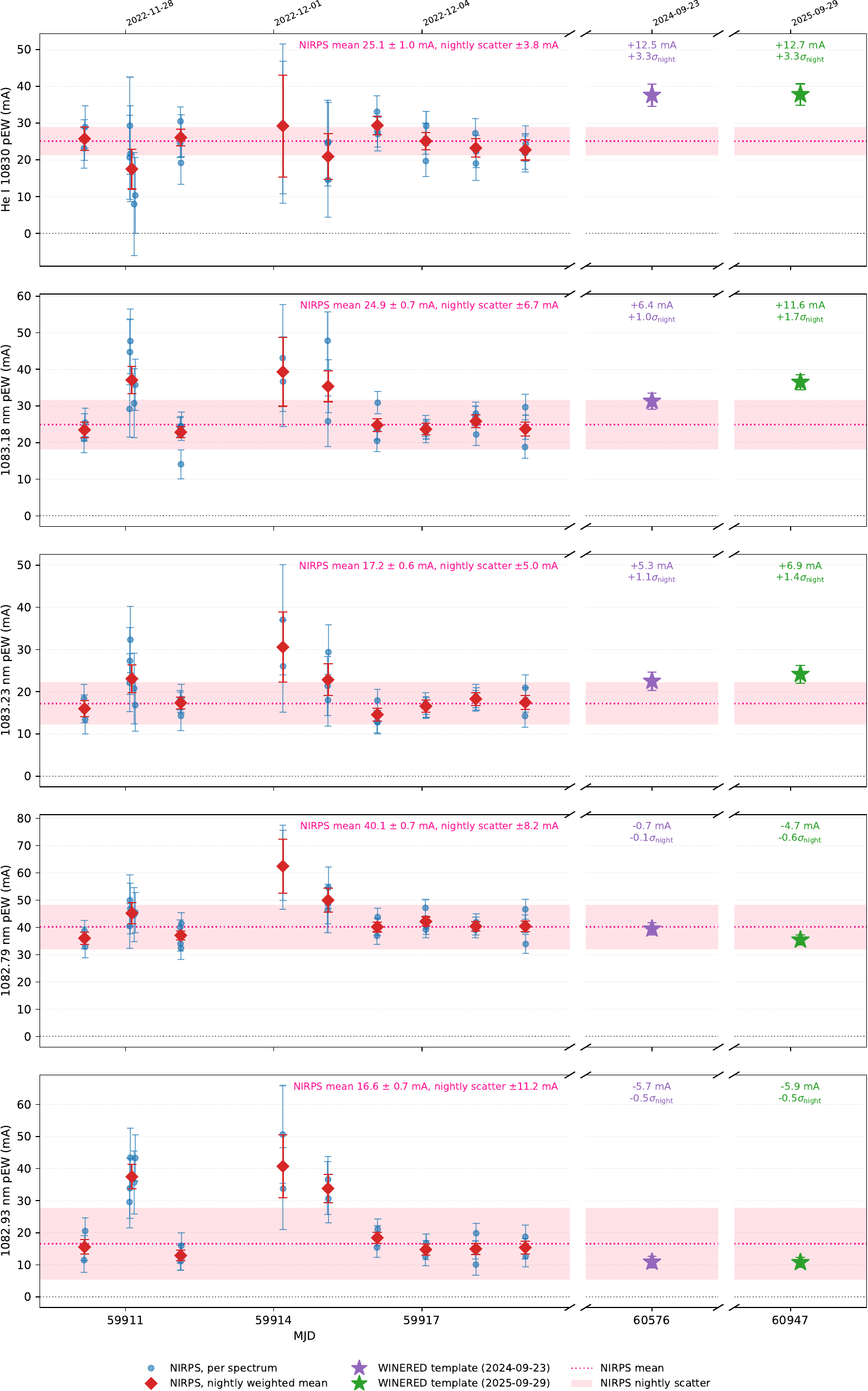}
    \caption{Idem to Figure~\ref{fig:heI_timeserie} adding four additional lines within the WINERED wavelength range. On average, there is no evidence for a systematic effect in which larger equivalent widths are measured for WINERED compared to NIRPS.   
    }
    \label{fig:HeI_NIRPS_WINRED_template_other_lines}
\end{figure*}

We also investigated potential systematic differences in equivalent-width measurements between the two instruments, specifically whether WINERED systematically yields larger equivalent widths than NIRPS. To this end, we measured four control lines at 1082.79, 1082.93, 1083.18, and 1083.23 nm using the same methodology applied to He\,I in both datasets. As shown in Figure~\ref{fig:HeI_NIRPS_WINRED_template_other_lines}, these measurements reveal no systematic tendency for WINERED to yield larger equivalent widths than NIRPS.

\section{He I Variability of M Dwarfs from the CARMENES Sample}
\label{app:carmenes_hei}

We sought to characterize the level of He\,I variability in mid-M dwarfs such as LHS\,1140. To do so, we revisited the CARMENES measurements of \citet{Fuhrmeister2020}, which provide multiepoch He\,I pseudo-equivalent widths (pEW) for a large sample of M dwarfs. We restricted the analysis to measurements classified as ``valid'' by \citet{Fuhrmeister2020}, i.e. stars for which the He\,I pEW could be measured reliably, and excluded known young stars in order to construct a cleaner comparison sample representative of field M dwarfs. We further retained only stars showing He\,I in absorption (positive pEW in our convention). The resulting sample is effectively limited to $T_{\rm eff}\gtrsim3300$~K and therefore contains no stars as cool as LHS\,1140 ($T_{\rm eff}\simeq3100$~K). Consequently, the CARMENES data do not provide a direct empirical measurement of the typical He\,I variability of an old mid-M dwarf with the temperature of LHS\,1140. They nevertheless allow us to examine systematic trends toward the cool end of the well-measured sample.

\begin{figure*}[htbp]
    \centering
    %\plottwo
    %      {HeIEW_vs_Teff_poyfit.pdf}
    %    {fractional_pEW_vs_Teff_polyfit.pdf}
    \includegraphics[height=5.5cm]{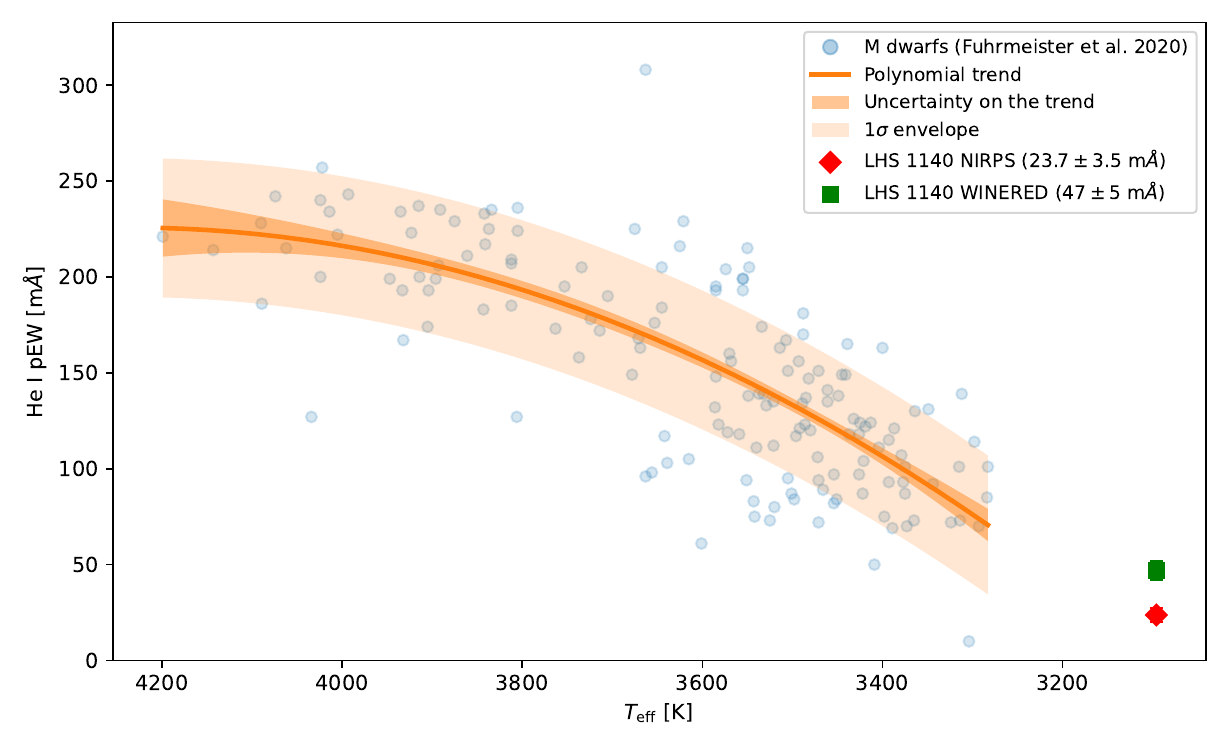}
    \hfill
    \includegraphics[height=5.5cm]{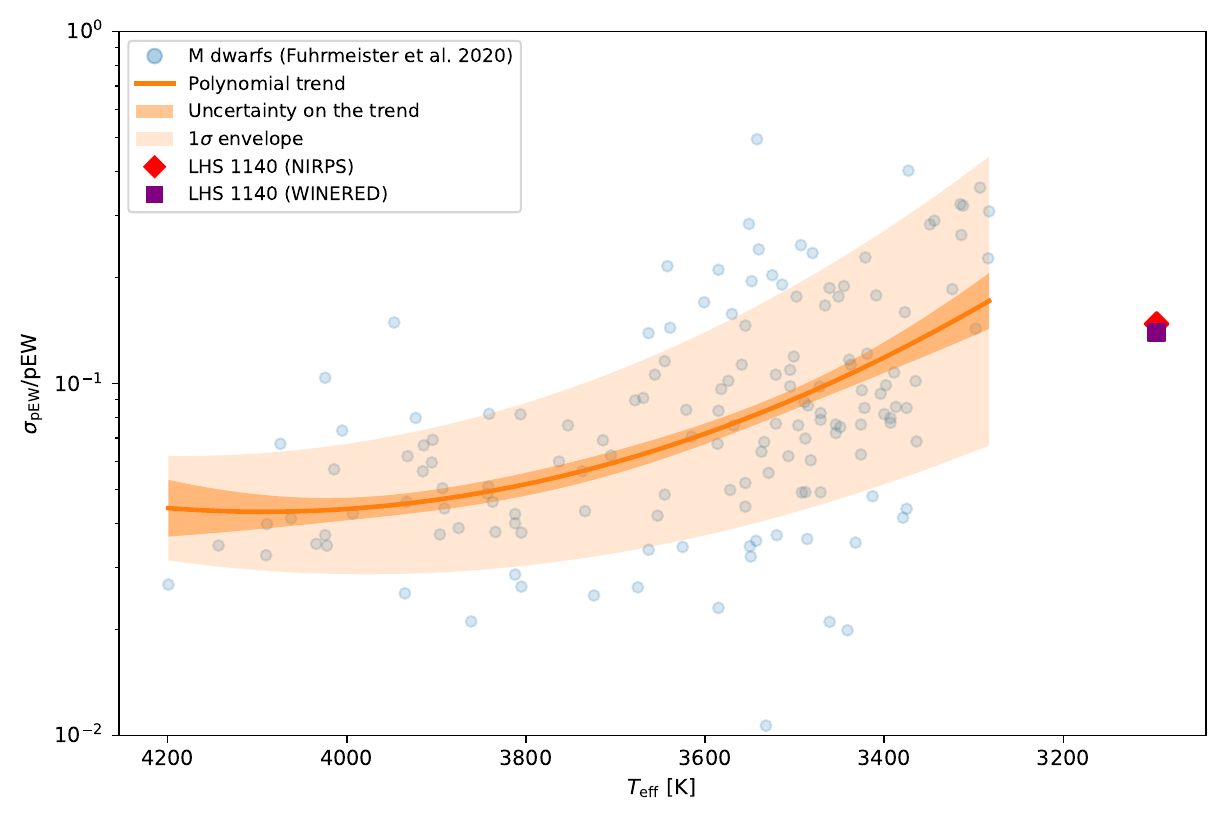}

      \caption{Left: He\,I 10830\,\AA\ pseudo-equivalent width (pEW) as a function of effective temperature for the CARMENES M-dwarf sample of \citet{Fuhrmeister2020}. The orange curve shows a polynomial fit to the pEW--$T_{\rm eff}$ relation; the darker shaded region indicates the uncertainty on the fitted trend, while the lighter region shows the $1\sigma$ dispersion of the stellar measurements about the trend. The LHS\,1140 measurements from NIRPS and WINERED (in transit during the 2024 event) are shown for comparison. Right: similar to (a) but showing the fractional He\,I 10830\,\AA\ variability, $\sigma_{\rm pEW}/\mathrm{pEW}$.}
    \label{fig:combined}
\end{figure*}

Figure~\ref{fig:combined} summarizes these trends. The left panel shows the mean He\,I pEW as a function of $T_{\rm eff}$. The CARMENES measurements display a clear decrease in He\,I absorption strength toward lower effective temperature. The polynomial trend and its dispersion illustrate that the line becomes progressively weaker toward the coolest stars for which reliable measurements are available. The NIRPS and WINERED measurements of LHS\,1140 are shown for comparison. Although LHS\,1140 lies beyond the cool edge of the ``valid'' CARMENES sample, its measured He\,I pEW is broadly consistent with an extrapolation of this decreasing trend.

The right panel of Figure~\ref{fig:combined} shows the fractional He\,I variability,
\begin{equation}
    f_{\rm var} =
    \frac{\sigma_{\rm pEW}}{\langle {\rm pEW}\rangle},
\end{equation}
as a function of $T_{\rm eff}$, where $\langle {\rm pEW}\rangle$ is the He\,I pEW measured from the median template spectrum of a given star, and $\sigma_{\rm pEW}$ is the standard deviation of its multiepoch pEW measurements inferred from the median absolute deviation reported by \citet{Fuhrmeister2020}. In contrast to the mean pEW, the fractional variability increases toward lower effective temperature. This trend is accompanied by substantial star-to-star dispersion, but indicates increasingly large fractional changes in the He\,I line toward the cool end of the CARMENES sample. The rise in fractional variability is at least partly driven by the weakening mean He\,I absorption, since a comparable absolute change in pEW represents a larger fraction of the line strength in cooler stars. Nevertheless, the CARMENES measurements provide no empirical indication that He\,I becomes intrinsically stable toward mid-M spectral types.

LHS\,1140 itself extends this behavior to lower temperature. The nine NIRPS epochs yield a weighted mean He\,I pEW of $23.55\pm1.07$~m\AA, whereas our independent reanalysis of the WINERED stellar templates gives $36.6\pm4.8$~m\AA\ in 2024 and $36.5\pm4.6$~m\AA\ in 2025. The two WINERED measurements are mutually consistent, but individually exceed the NIRPS mean at approximately $2.7\sigma$. The corresponding fractional variability inferred from the NIRPS measurements and from the combined NIRPS--WINERED baseline is shown in the right panel of Figure~\ref{fig:combined}. These measurements indicate that appreciable epoch-to-epoch variability of the disk-integrated stellar He\,I line is present in LHS\,1140 itself.

We therefore regard the CARMENES measurements as evidence that stellar He\,I variability remains relevant toward cool M dwarfs, while emphasizing that the lack of reliable CARMENES measurements below $\sim3300$~K prevents a direct empirical calibration at the temperature of LHS\,1140. Additional high-resolution, multiepoch spectroscopy of old mid-M dwarfs is required to establish the distribution of He\,I variability in this temperature regime.

%\begin{figure*}[htpb]
%    \centering
%        \includegraphics[width=\textwidth]{figures/HeIEW_vs_Teff_poyfit.pdf}
%    \caption{He\,I 10830\,\AA\ pseudo-equivalent width (pEW) as a function of effective temperature for the CARMENES M-dwarf sample of \citet{Fuhrmeister2020}. The orange curve shows a polynomial fit to the pEW--$T_{\rm eff}$ relation; the darker shaded region indicates the uncertainty on the fitted trend, while the lighter region shows the $1\sigma$ dispersion of the stellar measurements about the trend. The LHS\,1140 measurements from NIRPS and WINERED (in transit during the 2024 event) are shown for comparison.}
        
 %   \label{fig:heIEW_vs_Teff}
%\end{figure*}

%\begin{figure*}[htpb]
%    \centering
%        \includegraphics[width=\textwidth]{figures/fractional_pEW_vs_Teff_polyfit.pdf}
%        \caption{Fractional He\,I 10830\,\AA\ variability, $\sigma_{\rm pEW}/\mathrm{pEW}$, as a function of effective temperature for the CARMENES M-dwarf sample of \citet{Fuhrmeister2020}. Orange points show the mean fractional variability in 100-K bins, with the shaded region indicating the $1\sigma$ dispersion. The LHS\,1140 NIRPS and WINERED measurements follows the general trend of increased variability towards lower $T_{\rm eff}$.}
    \label{fig:HeI_fractional_variability}
%\end{figure*}

%\begin{figure}[htbp]
%   \centering
%    \subfloat[\label{fig:HeI_left}]{
%            \includegraphics[height=0.3\linewidth]
%            {figures/HeIEW_vs_Teff_poyfit.pdf} }
%        \hfill
%        \subfloat[\label{fig:HeI_right}]{
%            \includegraphics[height=0.3\linewidth]
%            {figures/fractional_pEW_vs_Teff_polyfit.pdf}
%        }

%    \begin{subfigure}[t]{0.48\textwidth}
    %    \centering
    %    \includegraphics[width=\linewidth]{figures/HeIEW_vs_Teff_poyfit.pdf}
    %    \caption{}
    %    \label{fig:HeI_left}
   %\end{subfigure}
%    \hfill
%    \begin{subfigure}[t]{0.48\textwidth}
%        \centering
%        \includegraphics[width=\linewidth]{figures/fractional_pEW_vs_Teff_polyfit.pdf}
%        \caption{}
%        \label{fig:HeI_right}
%    \end{subfigure}

%    \caption{Left: He\,I 10830\,\AA\ pseudo-equivalent width (pEW) as a function of effective temperature for the CARMENES M-dwarf sample of \citet{Fuhrmeister2020}. The orange curve shows a polynomial fit to the pEW--$T_{\rm eff}$ relation; the darker shaded region indicates the uncertainty on the fitted trend, while the lighter region shows the $1\sigma$ dispersion of the stellar measurements about the trend. The LHS\,1140 measurements from NIRPS and WINERED (in transit during the 2024 event) are shown for comparison. Right: similar to (a) but showing the fractional He\,I 10830\,\AA\ variability, $\sigma_{\rm pEW}/\mathrm{pEW}$.}
%    \label{fig:combined}
%\end{figure}

\end{appendix}

\clearpage

\bibliography{manuscript}{}
\bibliographystyle{aasjournal}



\end{document}